# Comparative Analysis and Evaluation of Two Prestressed Girder Bridges

**Mehrdad Aghagholizadeh[1*], Necati Catbas[2]**

[1] Lecturer, Southern Methodist University, Department of Civil and Environmental Engineering, Dallas, TX, USA
[2] Professor, University of Central Florida, Department of Civil, Environmental and Construction Engineering, Orlando, FL, USA

*** Correspondence:**
Mehrdad Aghagholizadeh
mehrdadag@smu.edu

**Keywords: Florida I-Beam, AASHTO Type Girder, Prestressed Concrete, Load Rating Factor, Bridge Reliability, Prestress Loss, Virtual Load Test.**

**Abstract**

A comparative analysis of two bridges constructed with the most commonly used girder types in Florida is carried out. The girder types that the bridges are employed for this study are AASHTO Type III (American Association of State Highway and Transportation Officials) and Florida I-Beam. Two bridges that have exactly the same specifications but with different girder type are analyzed under baseline state and different prestress loss cases. Finite element models of these bridges are utilized and the FE models are subjected to two types of virtual load tests of Florida legal loads namely C5 and SU4. Florida I-Beam tends to have higher load carrying capacity, higher lateral stiffness, cost efficiency and better element level reliability when compared to AASHTO Type girders.

## 1 Introduction

Bridges are important and inevitable components of transportation networks, designed and constructed with the aim of achieving safe, efficient, cost-efficient and time saving transportation. Throughout the years, in return for answering these aforementioned needs, bridge engineers have explored new design philosophies, high technology materials, real life issues and practices as well as the needs of general public. Type of the bridges may differ based on the material or design technique. Concrete or steel girder bridges could be given with respect to material difference whereas prestressed or posttensioned bridges might be related to design philosophy. Concrete girders allowed engineers to construct bridges for centuries whereas pre-stressed bridges rendered new designs possible with multiple and longer spans. Nowadays, pre-stressed concrete bridges constitute approximately 50% of newly built structures in US (PCA, 2004).

Use of pre-stressed members has many advantages due to their increased strength and durability. Excessive deflection of long spans is also a critical issue to be considered for such bridges. However, pre-stressed members not only provide reduction in deflection since they would already be bent the opposite way that bridge loads will mainly be applied, but also endure greater loads due to increased internal compression that will compensate a big portion of tension that will be induced by downward loads. Concrete girders together with pre-stressing application let clear majority of geometric forms be designed.

AASHTO (American Association of State Highway and Transportation Officials) I-beam and bulb T-beam have been widely used for concrete girders bridges across the US. Subsequently, ascent in traffic flow rate, need for reduced construction time and cost drove officials to create a more economic, efficient design. Consequently, Florida I-Beam (FIB) was created with the partnership of Florida Department of Transportation (FDOT) and academic researchers to meet these requirements (FDOT, 2009a). In comparison to AASHTO Type beams, FIBs show consistent and stabilized characteristics during site placement, easy manufacturing with only-web height adjustable forms and identical flange sizes. Besides, FIBs have larger sections both in web and flanges that allow higher number of strands to be placed in (up to 72 – 0.6 in diameter strands) and also higher strength (8 – 10 ksi), thus, carrying the same or higher magnitudes of loads with less number of beams and with higher clearance. This design eventually results in cost effective construction by saving of about 24%, which was given in a study by FDOT (FDOT, 2009a; 2009b). These findings and studies led FDOT to decide using FIBs instead of AASHTO beams (FDOT, 2009a).

Several studies were performed by FDOT on cost analysis of these two different bridges indicating FIBs are much more effective than AASHTO Type girders. In addition, one recent study outlined the development of 3D FE (finite element) models and their results for standard AASHTO based analysis and evaluation (Catbas et a., 2013). Another study stated the results of a comparative evaluation of the AASHTO Type III and FIB bridge (Peng and Catbas, 2014), in this study authors also considered additional permanent load other than dead loads. However, although there are some documentation in the literature about load carrying capacities from a comparison point of view, there is still need for deeper investigation on capacity, strength, reliability and durability to explore the new FIB designs compared to commonly used AASHTO Type girder bridges.

In this study, a comparative evaluation of two bridges, one built with AASHTO Type III girders and the other with FIBs, is carried out in terms of load rating factor and reliability indices under different structural conditions and virtual load testing scenarios. The first as-is condition represents the newly designed bridges and can be thought as the baseline case. Study on other cases gives information about how the load rating factors and reliability indices will change when there are different scenarios applied on the bridges such as pre-stress loss in both interior and exterior girders. In the emerge of studies on model updating to have reliable finite element model (Aghagholizadeh and Catbas 2015), this study will help to better understand how these property uncertainties effect the load carrying capacity of the structure using virtual load testing.

## 2 Bridge Specifications

Cross sections of the bridges with AASHTO Type III girder and FIB girders are presented (Figure 1). This section is comprised of six girders and the second bridge with the same load-carrying characteristics. The critical details for the appropriate FE modeling of the prestressed sections are presented with the necessary assumptions made for this study.

Two different models subject to this study are three span (90 ft each) bridges supported with three circular columns at the end of each section and with 41.5ft long beam cap on their top. Two cross sections that belong to two different models have the same section widths 43’-1” each but with different girder spacing that is 7'6" for AASHTO Type III and 12'6" for FIB as can be seen in Figure 1. Both types of girders have the same 45 in. depth. Each AASHTO Type III girders constitute 26-0.6 in low-relaxation prestressing strands whereas FIB girders contain 42- 0.6 in low-relaxation strands.

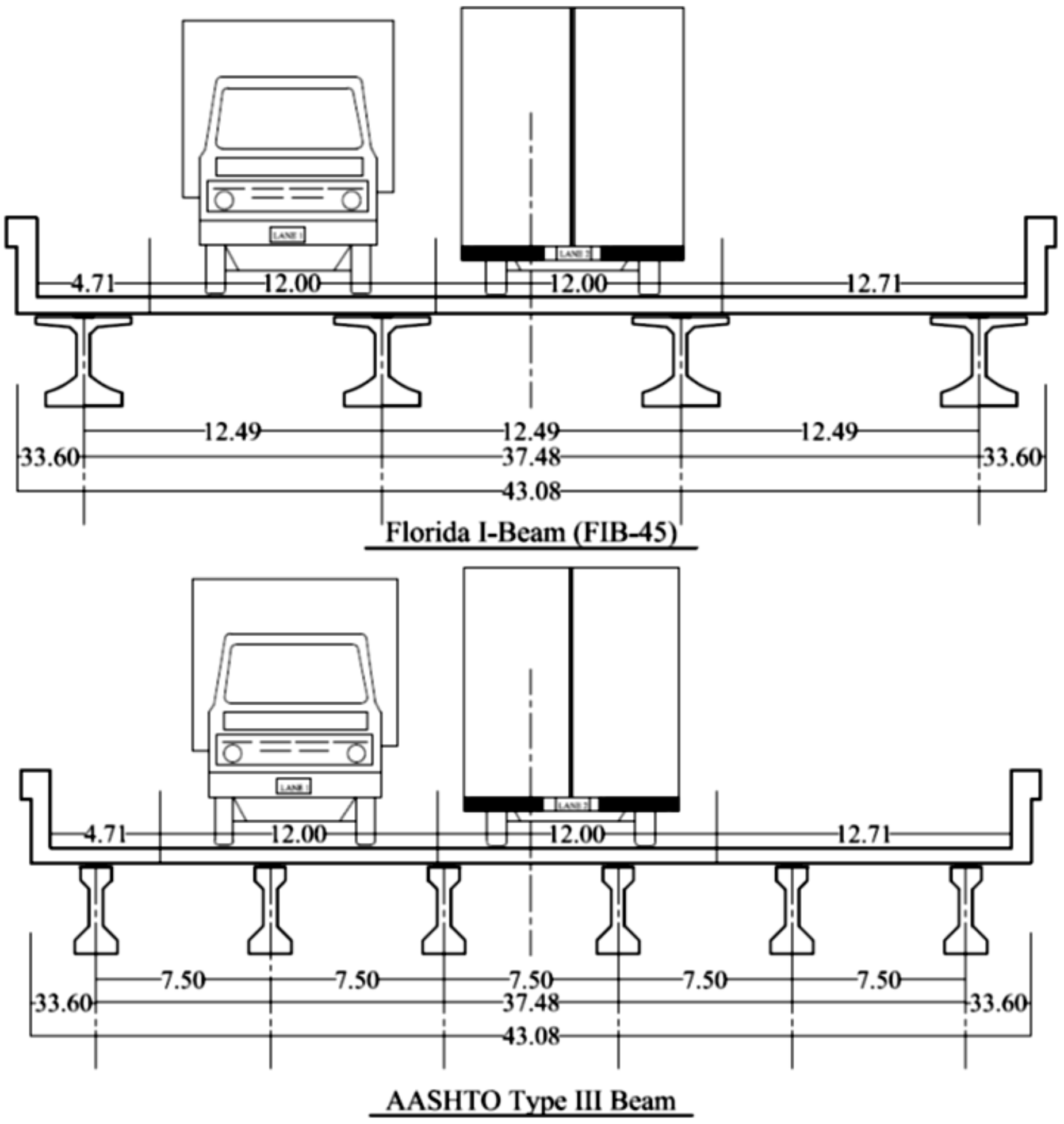

**Figure 1.** Cross Section of the bridges with Florida I beam (top), AASHTO Type III (bottom)

## 3 Method and Evaluation

### 3.1 Finite element model description

A finite element model using CSiBridge (ver. 15.2.0) is used to model and analyze the two types of Florida I-Beam and AASHTO Type prestressed concrete girder bridges (Figure 2). To define Florida I-Beam section, CSiBridge Section designer is used (Computers & Structures, Inc., 2017). Slab thickness is assumed to be 8 inches with 2 inches of haunch. The deck and columns concrete is cast in place (CIP) with compressive strength of 4 ksi. The same CIP concrete is used for abutments and beam caps. Girders are prefabricated and made with normal weight, 8.5 ksi compressive strength concrete. Cross diaphragms are used in every one third of the spans with depth and width of 19 in and 12 in respectively.

The deck is modeled using shell elements with six degrees of freedom at each node. Girders, columns and beam caps are modeled using frame elements in the software. The first 3-span bridge model is defined with 12 FIB-45 girders and 168 tendons, then the other mode with l8 AASHTO Type III girders and the total number of 156 tendons are defined for the entire bridge model (barker and Puckett, 1997; Nilson et al., 2010). The flexural capacity of the sections are computed and it is seen that the cross-section capacity of FIB bridge is 17% higher than AASTO Type III girder bridge. The cross-sectional properties are given in Table 1.

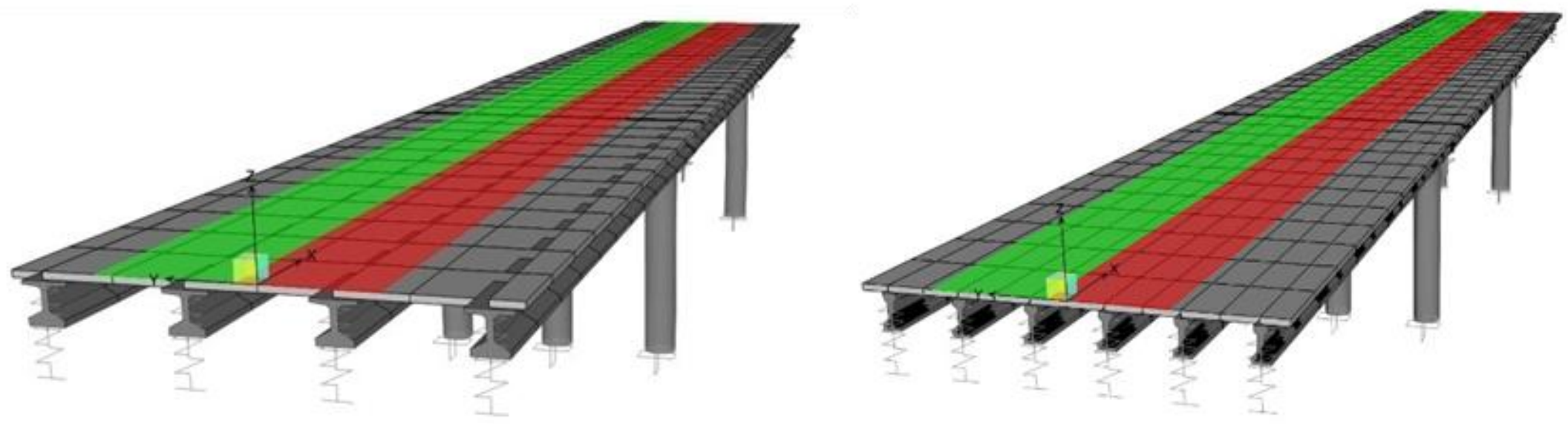

**Figure 2.** Finite Element Models: FIB (Left) AASHTO Type (Right)

For the FIB model, 12 post-tensioned girders are defined. Each girder reinforced with 42 0.6 inch low relaxation strands. Tendons are modeled as elements. On the other hand AASHTO Type III girder bridge is modeled with 18 girders for the entire bridge. There are 26 0.6 in low relaxation tendons, which are modeled as bar elements.

**Table 1.** Section Properties

| FIB - 45 section properties | | | |
|---|---|---|---|
| 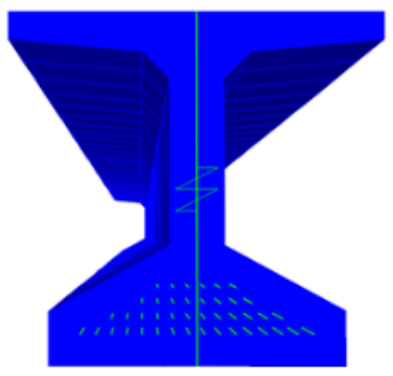 | 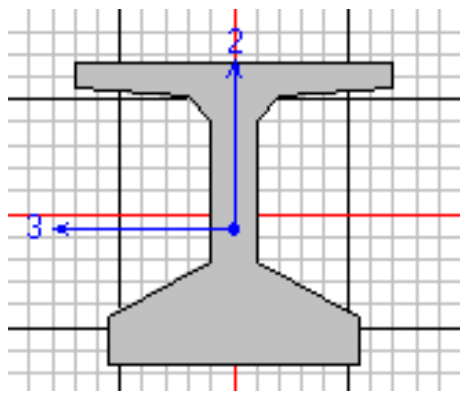 | Area | 869.58 in.$^2$ |
| | | $I_{xx}$ | 226,581 in.$^4$ |
| | | $I_{yy}$ | 81,327 in.$^4$ |
| | | $y_t$ | 24.79 in. |
| | | $y_b$ | 20.21 in. |
| AASHTO type III section properties | | | |
| 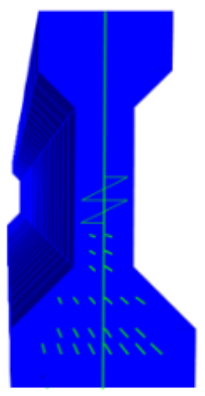 | 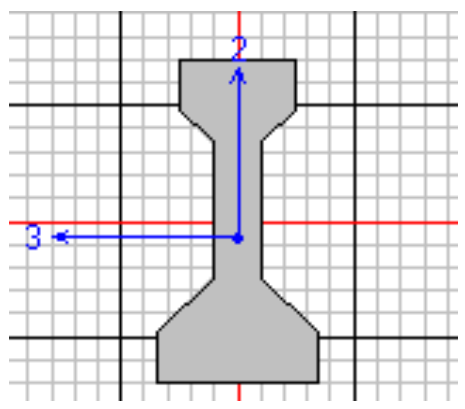 | Area | 559.5 in.$^2$ |
| | | $I_{xx}$ | 125,390 in.$^4$ |
| | | $I_{yy}$ | 12,217 in.4 |
| | | $y_t$ | 24.73 in. |
| | | $y_b$ | 20.21 in. |

### 3.2 Description of loads

Two different bridge models are analyzed under the Florida legal loads, C5 and SU4 that are given in the Florida Department of Transportation (FDOT) manual applied on the bridge (FDOT, 2010)

Florida legal loads are applied on the structure in compliance with the FDOT Bridge Load Rating Manual (FDOT, 2012). The same truck load is applied in each lane using only one truck per lane and they are not mixed. Additionally, dynamic load allowance is taken as IM=1.33 and no pedestrian live load is applied.

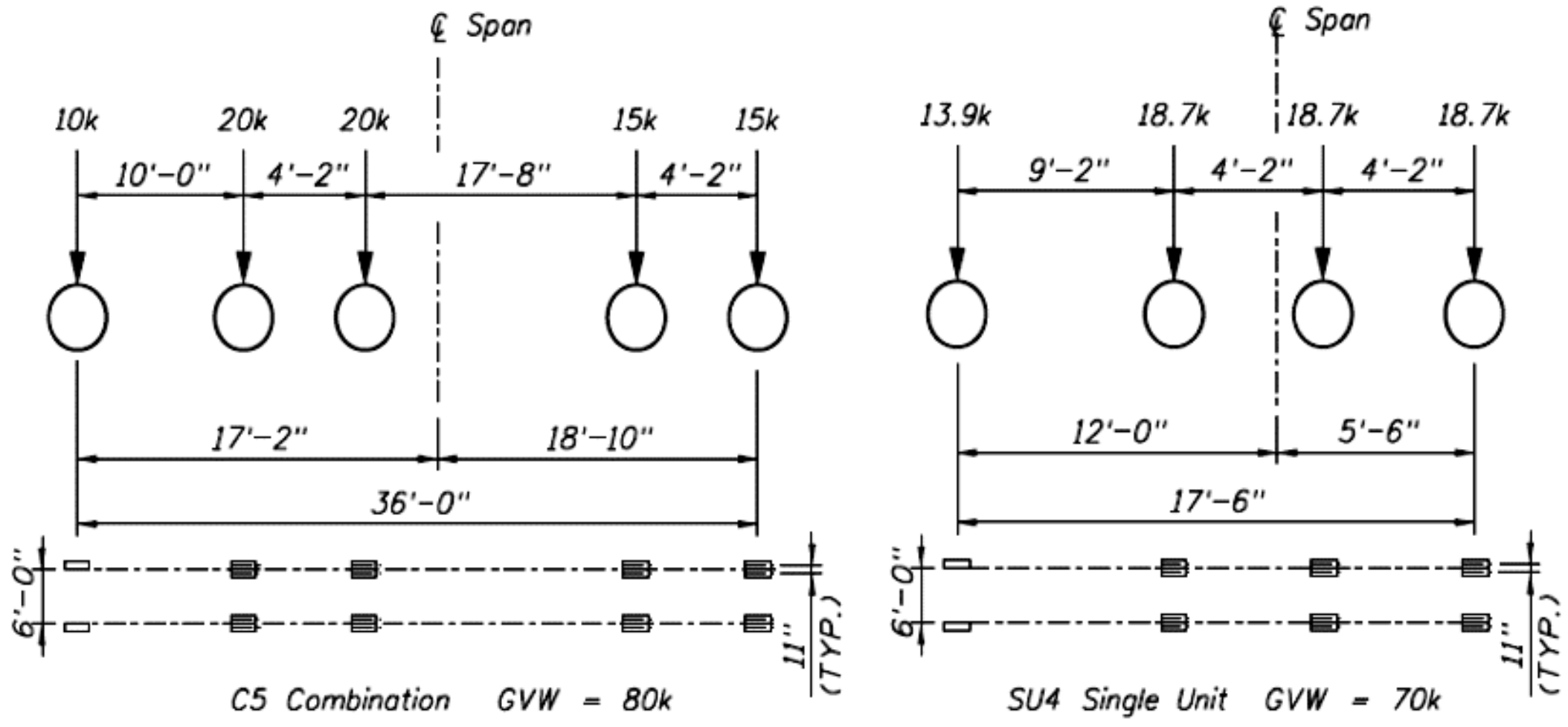

**Figure 3.** Florida Legal Loads that are used in FE Models

## 4 Description of Comparison Cases

The cases in question are most likely to be encountered in prestressed bridges both at the construction phase and during operation. AASHTO and FIB bridge models are analyzed under different cases namely baseline, immediate prestress losses and time dependent prestress loss. Variations in performance level of these bridges are presented in terms of load rating factors and reliability indices for the most critical sections in a comparative fashion.

### 4.1 Baseline case

Baseline case represents a completely healthy structure that is isolated from all possible external effects that might result in performance reduction. Any sort of immediate or time dependent losses together with other structural or environmental defects are ignored. A comparative study between two bridge models at their finest is carried out.

### 4.2 Prestress losses

Immediate prestress losses arise from the flaws that happen in either manufacturing or construction phase [8]. It can be broken down to different classification elements. In the scope of this study, elastic shortening of concrete, short term relaxation of steel and anchorage seating losses are taken into account as immediate prestress losses.

Elastic shortening occurs when concrete shortens and tendons accompany to this effect by losing some portion of their prestressing force after jacking. Knowing that $E_s$, $\varepsilon_{ES}$, $P_i$, $A_c$ and $E_c$ represent modulus of elasticity of steel, unit shortening in concrete, initial prestressing force, cross sectional area of concrete, modulus of elasticity of concrete and n is module of elasticity ratio of steel to concrete, elastic shortening can be defined with the following equation:

$$\Delta f_{pES} = E_s \varepsilon_{ES} = \frac{E_S P_i}{A_c E_c} = \frac{nP_i}{A_c} = nf_{cs} \qquad (1)$$

where $\Delta f_{pES}$ is the pre-stress loss due to concrete shortenting. Equation 1 expresses the loss in terms of the stress in the concrete $f_{cs}$, when the strand is located directly at the centroid of the cross section. If there exists a tendon eccentricity $e$ at the beam, the equation becomes:

$$f_{cs} = -\frac{P_i}{A_c}\left(1+\frac{e^2}{r^2}\right)+\frac{M_D e}{I_c} \tag{2}$$

where $r^2, M_D$ and $I_C$ represent radius of gyration, moment generated by the member's dead weight and moment of inertia respectively. For post-tensioned beams, $N$ representing the number of strands or strands pairwise, the following equation is suitable to find the loss of prestress in case consecutive jacking stages are applied on the element.

$$\Delta f_{pES} = \frac{1}{N}\sum_{1}^{N}\left(\Delta f_{pES}\right)_j \tag{3}$$

Steel relaxation is one of the common loss cases that can be either considered for immediate or time dependent based on duration taken into account. This type of loss can be determined based on the ratio $f_{pi}/f_{py}$ of the initial prestress to the yield strength of the reinforcement. The choice of $f_{pi}$ and $f_{py}$ for calculation purposes is restricted to certain limitations by the ACI 318-05 Code. In this study, $0.75\ f_{pu}$ is chosen as the upper limit for both initial prestressing force and the yield strength of the reinforcement. Subsequently, stress relaxation loss can be computed for any desired time interval through the following equation:

$$\Delta f_{pR} = f_{pi}'\left(\frac{\log t_2 - \log t_1}{10}\right)\left(\frac{f_{pi}'}{f_{py}} - 0.55\right) \tag{4}$$

where $f_{pi}'$ is the initial stress in steel to which the concrete element is subjected; $t_1$ and $t_2$ are time steps that represent jacking time in hours and desired final loss stage to be found respectively. For the calculation of immediate prestress loss, 18 hours and for long-term losses 1 year (8760 hours) is considered.

Another immediate type of loss considered is anchorage seating loss that is caused by the modification of tendons to their new state when the prestressing force is transferred from jacking to casting beds. Knowing that $\Delta_A$is the magnitude of slip, $L$ is the tendon length and $E_{PS}$ is the modulus of elasticity of the steel strands, anchorage seating loss can be given as:

$$\Delta f_{pA} = \frac{\Delta_A}{L}E_{ps} \tag{5}$$

Behavior of bridges may show differences during their operational life as they undergo time dependent effects. Although these effects enclose uncertainties that prevent exact determination particularly, it is possible to calculate time dependent prestress losses for the prestressed concrete case via empirical relations derived from codes of practice. In the scope of this study, long term relaxation of steel, creep loss and shrinkage loss are taken into account as time dependent prestress losses employing code defined formulations.

Determination of creep loss involves taking into consideration of different effects such as the amount of the applied load and its continuance, certain characteristics of concrete mixture that the prestressed element is made of, curing conditions, how old is the element when it is first loaded and ambient effects on the element. Based on the approximation that the relation between stress-strain and creep is linear, an empirical equation from ACI-ASCE committee can be used for proper estimation (ACI – ASCE Joint Committee, 1957):

$$\Delta f_{pCR} = K_{CR} \frac{E_{ps}}{E_c} \left( \bar{f}_{cs} - \bar{f}_{csd} \right) \tag{6}$$

where $K_{CR} = 2.0$ for pretensioned members, $\bar{f}_{cs}$ is concrete stress right after prestress transfer and $\bar{f}_{csd}$ is concrete stress after prestressing is carried out when all dead loads are subjected on the system.

In addition to the same effects that are considered in creep loss case, shrinkage loss incorporates dimensions of the member, type of the concrete mixture contents etc. According to Prestressed Concrete Institute (PCI), prestressing loss due to shrinkage can be estimated through the following equation:

$$\Delta f_{pSH} = 8.2 \times 10^{-6} \, K_{SH} E_{ps} \left( 1 - 0.06 \frac{V}{S} \right) (100 - RH) \tag{7}$$

where RH represents relative humidity, which is taken as 75% in this study and $K_{SH}$ coefficient is assumed the concrete is moist cured in 7 days and since the members are post-tensioned $K_{SH}$ is 0.77 (Preston, 1975; PCI 1999) .

All the losses that are calculated with the help of given equations are tabulated in Table 2.

**Table 2.** Loss Calculation Results

| Beam type | Type of losses | Loss stages | | Stress (psi) | Percent |
|---|---|---|---|---|---|
| Florida I beam | Initial losses | After tensioning ($0.75f_{pu}$) | (a) | 202500 | 100.0 |
| | | Elastic shortening $\Delta f_{ES}$ | (b) | -7540 | -3.7 |
| | | Steel relaxation $\Delta f_{PR}$ | (c) | -6447 | -3.2 |
| | | Anchorage seating $\Delta f_{PA}$ | (d) | -7812 | -3.9 |
| | | Total immediate loss | (b+c+d) | -21799 | -10.8 |
| | | *Final net stress* | (a-b-c-d) = e | *180701* | *89.2* |
| | Time dependent losses | Steel Relaxation $\Delta f_{pT}$ | (f) | -11433 | -5.7 |
| | | Creep loss $\Delta f_{pCR}$ | (g) | -4272 | -2.1 |
| | | Shrinkage $\Delta f_{pSH}$ | (h) | -3412 | -1.7 |
| | | Total time dependent losse | (f+g+h) | -19117 | -9.4 |
| | | *Final net stress* | (e-f-g-h) | *161584* | *79.8* |
| AASHTO Type III beam | Initial losses | After tensioning ($0.75f_{pu}$) | (a) | 202500 | 100.0 |
| | | Elastic shortening $\Delta f_{ES}$ | (b) | -7770 | -3.8 |
| | | Steel relaxation $\Delta f_{PR}$ | (c) | -6100 | -3. |
| | | Anchorage seating $\Delta f_{PA}$ | (d) | -8100 | -4 |
| | | Total immediate loss | (b+c+d) | -21970 | -10.8 |
| | | *Final net stress* | (a-b-c-d) = e | *180530* | *89.2* |
| | Time dependent losses | Steel Relaxation $\Delta f_{pT}$ | (f) | -11396 | -5.6 |
| | | Creep loss $\Delta f_{pCR}$ | (g) | -7204 | -3.6 |
| | | Shrinkage $\Delta f_{pSH}$ | (h) | -3698 | -1.8 |
| | | Total time dependent losse | (f+g+h) | -22298 | -11 |
| | | *Final net stress* | (e-f-g-h) | *158232* | *78.1* |

## 5 Description of Comparison Criteria

The two FIB and AASHTO Type girder bridges of this study is compared in terms of Load Rating Factor (LRF) and the Reliability Indices (RI). For the LRF and RI calculations, the most critical moment and capacity are selected for the inerior and exterior girders from all the girders of the entire bridge. The results reported for the interior and exterior girders for the critical sections.

### 5.1 Load rating factor (LRF)

"The load rating process is a component of the inspection process and consists of determining the safe load carrying capacity of structures, determining if specific legal or overweight vehicles can safely cross the structure and determining if structure needs to be restricted and the level of posting required" (FDOT, 2010).

$$RF = \frac{\varphi M_n - \gamma_{DC} M_{DC} - \gamma_{DW} M_{DW}}{\gamma_L M_{LL+IM}} \qquad (8)$$

where $RF$ = Rating factor; $M_n$ = Nominal moment resistance; $\varphi$ = Resistance factor for flexure; $\gamma_{DC}\ M_{DC}$ = Factored moment due to dead load of structural components and attachments; $\gamma_{DW}\ M_{DW}$ = Factored moment demand due to dead load of wearing surface and utilities; and $\gamma_L\ M_{LL+IM}$= Factored moment due to live load

In this paper, flexure load rating factor for Strength I Limit State is calculated for the most critical moment at the exterior and interior girders. In this limit state, the resistance for flexure $\phi$ is 1.0, the dead load moment factor $\gamma_{DC}$=1.25, future wearing surface moment factor $\gamma_{DW} = 1.25$ and live load moment factor $\gamma_L = 1.35$ considering 33% of dynamic allowance (IM) in the live load analysis.

### 5.2 Reliability index (RI)

Current AASHTO LRFD Bridge design code is based on reliability analysis procedure (Nowak, 1995; Ghosn and Frangopol, 1999). Performance of the structures is determined by means of load and resistance factors determined from the probability of failure and the reliability. The alternative way to express probability of failure is using the reliability index $\beta$. In normally distributed random variables the relation between probability of failure and the reliability index is as follows, $P_f = \Phi_{(-\beta)}$ and for the normally distributed cases this expression is exact (Nowak amd Collins, 2000; Shmerling and Catbas, 2009).

$$\beta = \frac{a_0 + \sum_{i=1}^{a}\left(a_i \mu_{X_i}\right)}{\sqrt{\sum_{i=1}^{a}\left(a_i \sigma_{X_i}\right)^2}} \qquad (9)$$

For the linear limit state function of the form,

$$g\left(X_1, X_2, ..., X_n\right) = a_0 + a_1 X_1 + a_2 X_2 + ... + a_n X_n \qquad (10)$$

In this study, member reliability indices are computed. In order to calculate reliability index, f section resistance moment, dead load moment, wearing surface moment and live load moments at the most critical section are obtained from the finite element model. Then using Tables 3 and 4, bias factor (λ) and coefficient of variation (V) from AASHTO code is selected. Afterwards from equation 11 and 12, mean value (μ) and standard deviation (σ) are calculated (AASHTO, 2003).

$$\mu_i = M_i \lambda_i \quad (11)$$

$$\sigma_i = V_i \mu_i \quad (12)$$

where *i* represents resistance, dead, wearing surface and live load moments. Considering AASHTO LRFD limit state function which is developed in terms of resistance and load effects for the strength limit state, the following formula is achieved in order to calculate reliability index.

Reliability Index for AASHTO Strength Limit State I:

$$\beta = \frac{\mu_R - \mu_{DL} - \mu_L}{\sqrt{\sigma_R^2 + \sigma_{DL}^2 + \sigma_L^2}} \quad (13)$$

where $\mu_R$ is the mean resistance, $\mu_{DL}$ is mean dead load, $\mu_L$ is mean live load, whereas $\sigma_R, \sigma_{DL}$ and $\sigma_L$ represents the standard deviations. Table 3 and 4 illustrates the statistical parameters used for bridge loading and resistance.

**Table 3.** Statistical values for Bridge load components

| **Load component** | **Bias ($\lambda_Q$)** | **COV ($V_Q$)** |
|---|---|---|
| Dead load: | | |
| Factory made | 1.03 | 0.08 |
| Cast in place | 1.05 | 0.10 |
| Asphalt wearing surface | 1.00 | 0.25 |
| Live load (with dynamic load allowance) | 1.10-1.20 | 0.18 |

**Table 4.** Statistical parameters of resistance for selected bridges

| Type of structure | Bias ($\lambda_R$) | COV ($V_R$) |
|---|---|---|
| Non-composite steel girders: | | |
| Moment (compact) | 1.12 | 0.10 |
| Moment (noncompact) | 1.12 | 0.10 |
| Shear | 1.14 | 0.105 |
| Composite steel girders: | | |
| Moment | 1.12 | 0.10 |
| Shear | 1.14 | 0.105 |
| Reinforced concrete T-beams: | | |
| Moment | 1.14 | 0.13 |
| Shear w/steel | 1.20 | 0.155 |
| Shear w/o steel | 1.40 | 0.17 |
| Prestressed concrete girders: | | |
| Moment | 1.05 | 0.075 |
| Shear w/ steel | 1.15 | 0.14 |

## 6 Results and Discussions

The first analysis is to determine the flexural load rating factor for the baseline state, which is described as the perfect condition of the bridge isolated from all possible losses and environmental effects. The analysis is carried out for both bridge types and LRFs are calculated for each individual girder at different locations. The variation of LRF at 7.5 feet intervals are calculated and results are illustrated in the following Figure 4 to 7 for the baseline case.

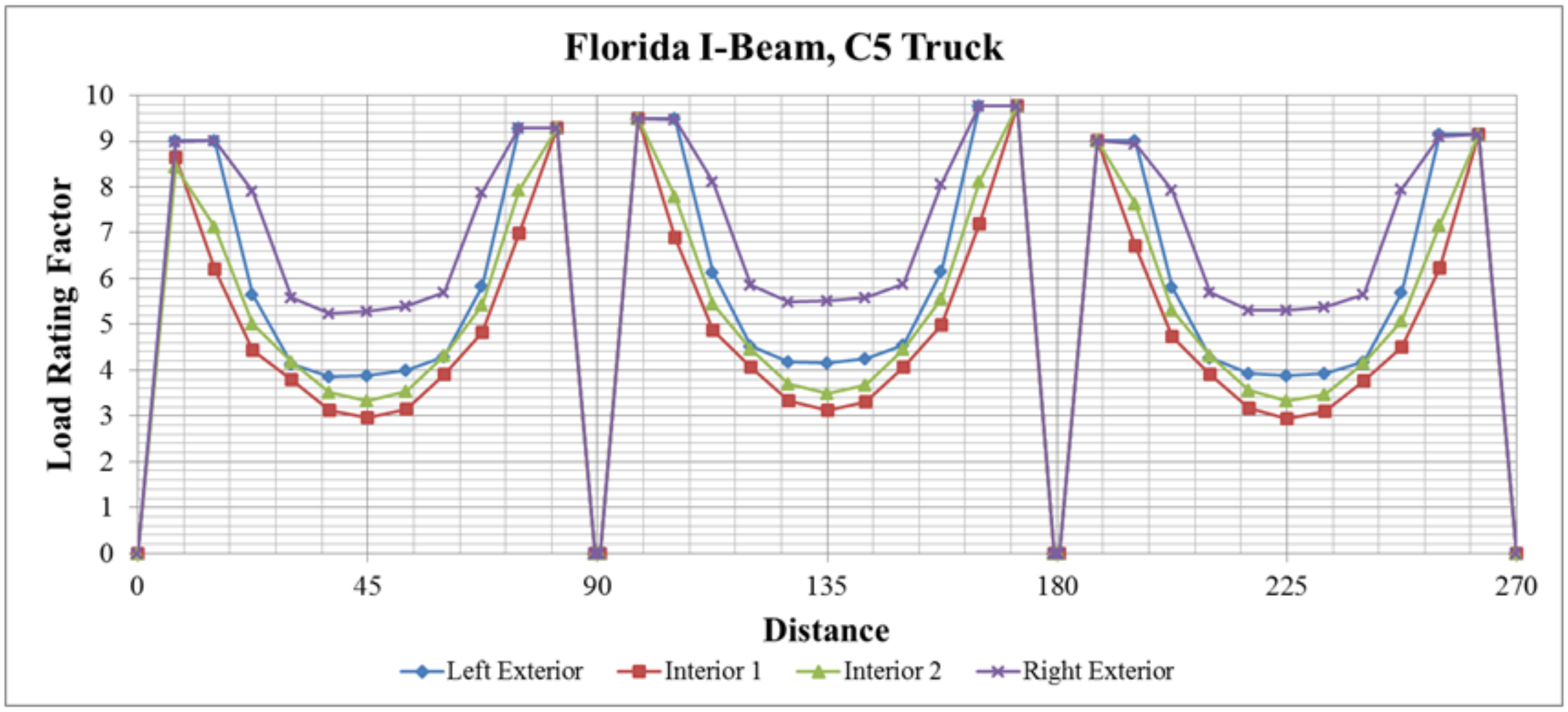

**Figure 4.** Load rating through Entire FIB bridge under C5 truck load

In Figure 4 and Figure 5, load rating factors for Florida I-Beam show very similar tendency and values for C5 and SU4 legal loads. Computed rating factors are lower at the interior girders and take higher values at the exteriors, which makes sense due to the placement of trucks and distribution of load over the girders. Additionally, right exterior girder seems to have a higher LRF than that of the left exterior which is due to the emergency lane is carried by the right exterior girder.

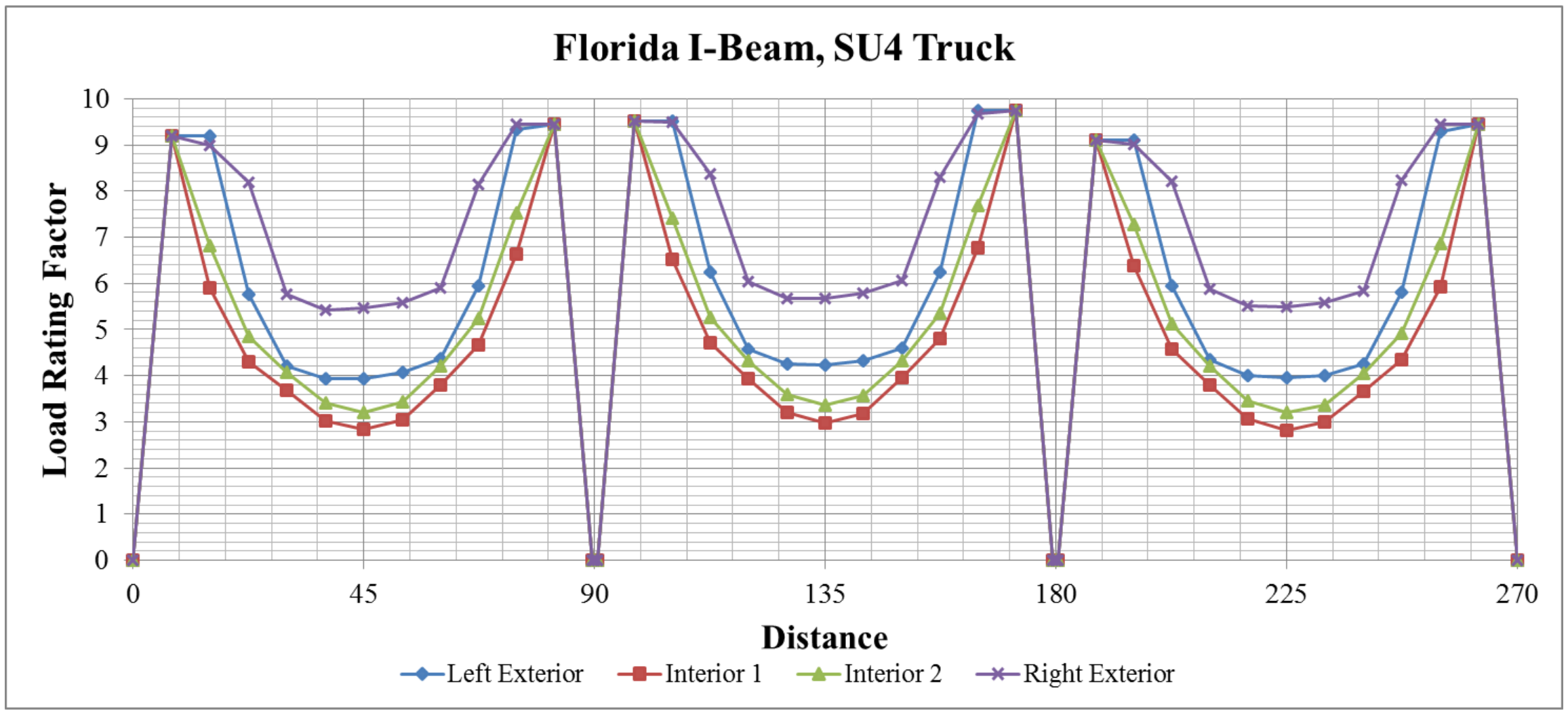

**Figure 5.** Load Rating Through Entire FIB Bridge under SU4 Truck Load

In Figure 6 and 7, LRFs at every 7.5 feet intervals are given for the baseline case along the entire bridge with AASHTO type III girder. The same tendencies explained for Florida I-beam is also observed for this beam with some minor differences. Interior girders again take the lowest LRF values; however, the magnitudes of the AASHTO Girder LRFs are lower at each span compared to Florida I-beam.

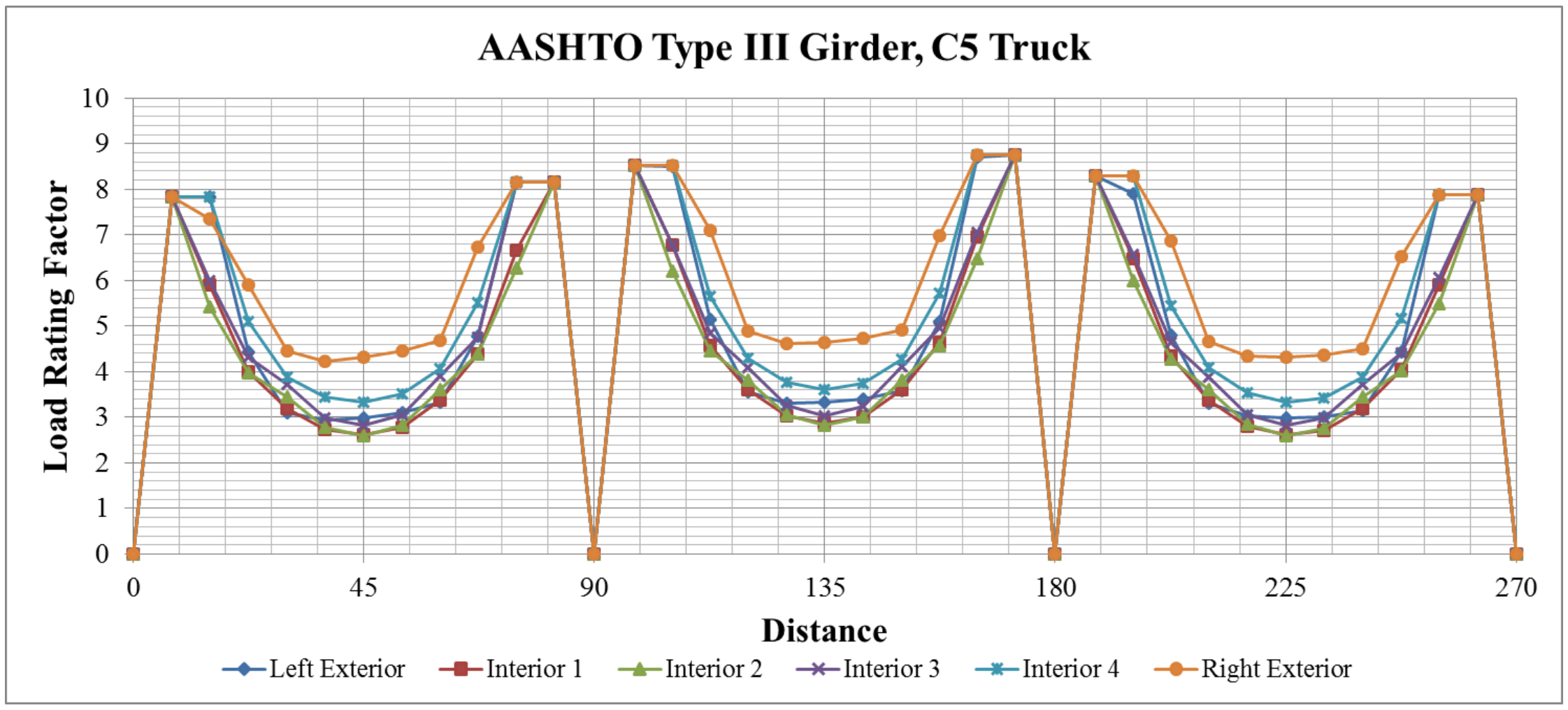

**Figure 6.** Load Rating Through Entire AASHTO Type Bridge under C5 Truck Load

To have a better understanding of LRF differences of AASHTO Type and FIB girders, the difference in LRF of AASHTO and FIB, with respect to AASHTO are presented. For the baseline case, this difference is 29.8% for exterior girders and 13.1% for interior girders under C5 truck load. Similarly, 28.3% for exterior girders and 14.6% for interior girders under SU4 truck load. These findings verify that the load carrying capacity of Florida I beam is higher than that of the AASHTO type III beam as pointed out by the FDOT. In other words, FIB load rating factor of 3.88 for C5 translates to 310.4 kips (3.88 x 80 kips), whereas AASHTO 2.99 load rating translates to 239.2 kips (29.8% difference between 310.4 kips 239.2 kips). Similarly, FIB load rating of 2.94 translates to 235.2 kips (2.94 x 80 kips), whereas 2.60 load rating factor for AASHTO for interior girder gives 208 kips (13% difference between 235.2 kips and 208 kips). The possible live load that can be carried out under different cases can be determined in a similar fashion.

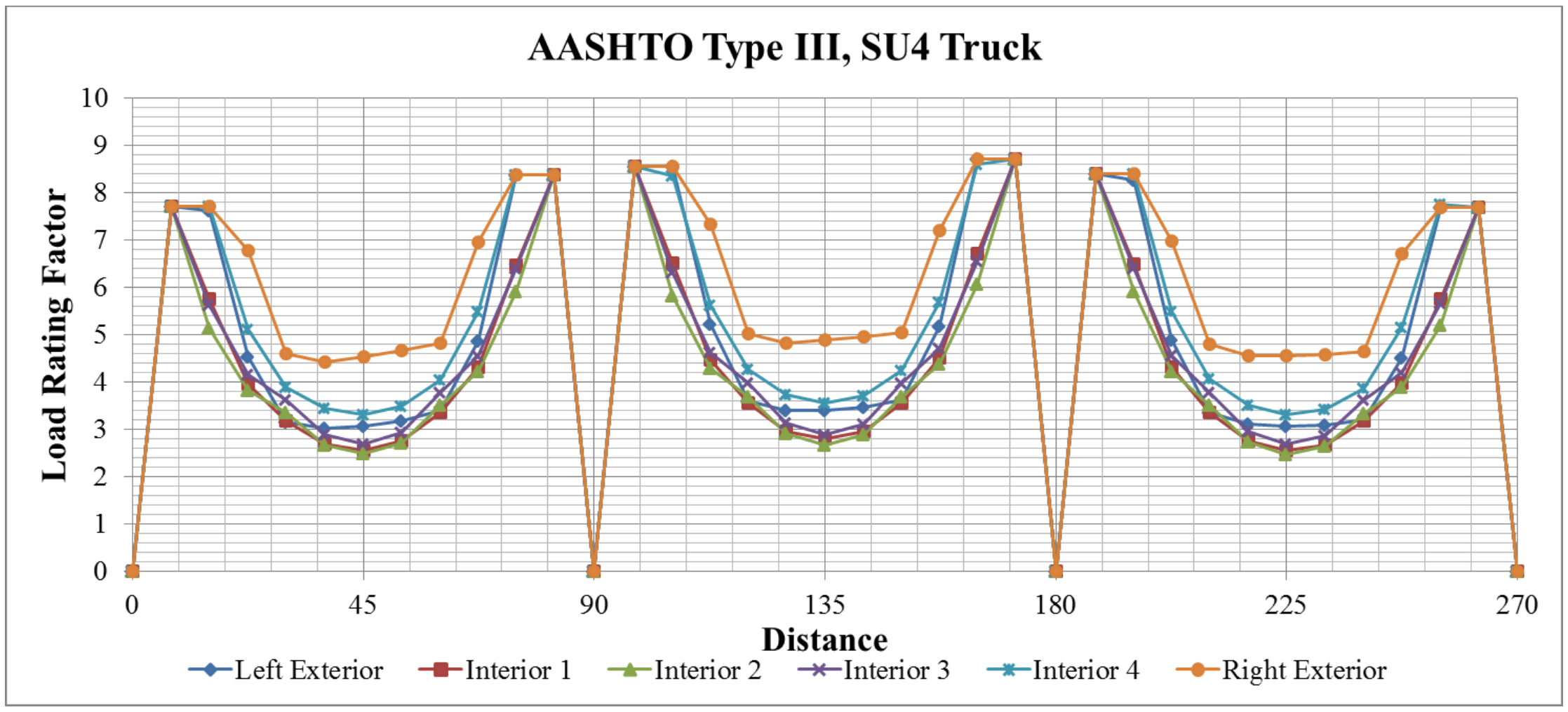

**Figure 7.** Load Rating Through Entire AASHTO Type Bridge under SU4 Truck Load

From the detailed evaluation and assessment of load rating factors for the baseline case and considering each span is simply supported for both bridges, it can be concluded that the load rating factor takes its lowest value at the midspan in each girder and that the minimum load rating value is chosen as the governing critical value since the first failure is expected at this member and it is directly related to overall performance of the structure. In Tables 5 and 6, critical values of LRFs for each structural condition are tabulated for the interior and exterior girders. The differences of each condition with respect to baseline case are also included in these tables. As the losses increase cumulatively, load rating factors, or load carrying capacity in other words decrease in magnitude causing the difference to go higher in percentage.

**Table 5.** Florida I-Beam analysis results

| Analysis | Girder location | Vehicle type | Baseline | Immediate losses | Change wrt Baseline (%) | Long term losses | Change wrt Baseline (%) |
|---|---|---|---|---|---|---|---|
| Load Rating Factor | Exterior | C5 | 3.88 | 3.32 | 14.53 | 3.02 | 22.19 |
| | Interior | | 2.94 | 2.47 | 15.90 | 2.22 | 24.37 |
| | Exterior | SU4 | 3.94 | 3.37 | 14.37 | 3.07 | 22.05 |
| | Interior | | 2.83 | 2.37 | 16.10 | 2.14 | 24.54 |
| Reliability index β | Exterior | C5 | 5.97 | 5.27 | 11.77 | 4.85 | 18.81 |
| | Interior | | 4.91 | 4.14 | 15.63 | 3.69 | 24.91 |
| | Exterior | SU4 | 6.00 | 5.30 | 11.64 | 4.88 | 18.63 |
| | Interior | | 4.82 | 4.05 | 16.07 | 3.59 | 25.57 |

In Table 5 and Table 6, reliability indices are also included. These reliability indices are tabulated with the same procedure for the most critical location as summarized for the LRF calculations. For each member in every span, reliability indices are calculated using the output of the most critical flexural moment values. The same trend is seen as in LRF.

**Table 6.** AASHTO type III beam results

| Analysis | Girder location | Vehicle type | Baseline | Immediate losses | Change wrt Baseline (%) | Long term losses | Change wrt Baseline (%) |
|---|---|---|---|---|---|---|---|
| Load Rating Factor | Exterior | C5 | 2.99 | 2.54 | 15.02 | 2.28 | 23.72 |
| | Interior | | 2.60 | 2.20 | 15.27 | 1.97 | 24.17 |
| | Exterior | SU4 | 3.07 | 2.61 | 14.98 | 2.34 | 23.68 |
| | Interior | | 2.47 | 2.10 | 15.15 | 1.88 | 24.06 |
| Reliability index β | Exterior | C5 | 4.88 | 4.18 | 14.42 | 3.72 | 24.23 |
| | Interior | | 4.40 | 3.68 | 16.36 | 3.21 | 26.95 |
| | Exterior | SU4 | 4.93 | 4.23 | 14.12 | 3.78 | 23.31 |
| | Interior | | 4.28 | 3.56 | 16.93 | 3.09 | 27.87 |

Starting from the baseline case and going through prestress losses one by one, reduction in LRF and reliability index magnitude can be seen clearly. It is obvious to see that there are direct correlations between the variation of load rating factors and reliability indices. It is seen that the long-term losses have the most impact in terms of load rating and reliability for the most critical locations. While the load carrying capacity is reduced, it is also seen that the element reliability at the critical sections is reduced, thereby increasing the probability of failure.

## 7 Conclusion

A comparative study between two bridges constructed with the most commonly used girders in Florida, AASHTO Type III girder and Florida I-Beam, is carried out. Florida I-Beams tend to have higher load carrying capacity, higher lateral stiffness, cost efficiency and better reliability when compared to AASHTO Type girders. Two bridge structures have the same loading and deck geometry specifications. The different girder type bridges are analyzed under both baseline and different performance loss states. In this study, finite element models of these bridges are developed using commercial software and the bridges are subjected to one lane loading with Florida legal loads, namely C5 and SU4. Load rating factors are calculated based on the results acquired from different cases and reliability of critical members are investigated.

From the results, it can be concluded that for all the cases that were investigated, load rating factors are always higher for bridges constructed with FIBs than that of bridges with AASHTO girders. This provides additional live load carrying capacity for the bridge analyzed with FIB girders.

The load rating analysis for the C5 Truck (80 kips) presents different load carrying capability. FIB bridge has a load rating factor of 3.88 for the critical exterior girder, meaning 310.4 kips can be carried out safely, whereas AASHTO bridge load rating factor is 2.99, meaning 239.2 kips. That means there is 71 kips difference between 310.4 kips 239.2 kips. Similarly, FIB bridge load rating of 2.94 translates to 235.2 kips, whereas AASHTO bridge 2.60 load rating factor for for interior girder gives 208 kips. That means there is 27 kips difference between 235.2 kips and 208 kips.

When the reliability indices are evaluated for the most critical interior girder under C5 loading, it is seen that the FIB and AASHTO bridges has reliability indices of 3.69 and 3.21, respectively. These indices can be given as probability of failure with 1.12E-4 and 6.64E-4 for the FIB and AASHTO bridges respectively. From this analysis, it can be concluded that the probability of failure of the interior bridge girder of AASHTO is approximately 6 times the FIB bridge girder. It is also important to note that system-level reliability of the AASHTO bridges can be expected to be higher due to more girders placed in parallel. This study will be in future papers.

## Conflict of interest

The authors declare no potential conflict of interests.